\documentstyle[twocolumn,seceq]{jpsj}

\newcommand{\be}{\begin{equation}}
\newcommand{\ee}{\end{equation}}
\newcommand{\bea}{\begin{eqnarray}}
\newcommand{\eea}{\end{eqnarray}}
\newcommand{\lsim}{\raise.35ex\hbox{$<$}\kern-0.75em\lower.5ex\hbox{$\sim$}}
\newcommand{\gsim}{\raise.35ex\hbox{$>$}\kern-0.75em\lower.5ex\hbox{$\sim$}}
\def\dfrac#1#2{{\displaystyle\frac{#1}{#2}}}
\def\cfrac#1#2{\dfrac{\mathstrut #1}{#2}}

\def\runtitle{Dynamical Structure Factors of the Spin-1/2 $XXZ$ Chain
}
\def\runauthor{Yasuhiro {\sc Saiga} and Yoshio {\sc Kuramoto}}

\title
{
Dynamical Structure Factors of the Spin-1/2 $XXZ$ Chain \\
with Inverse-Square Exchange and Ising Anisotropy
}

\author
{ 
Yasuhiro {\sc Saiga}\footnote{E-mail: saiga@cmpt01.phys.tohoku.ac.jp} and Yoshio {\sc Kuramoto}
}

\inst
{
Department of Physics, Tohoku University, Sendai 980-77
}

\recdate
{\hspace{2cm}
}

\abst
{
The dynamical properties of the $S=1/2$ antiferromagnetic $XXZ$ chain are studied by the exact diagonalization and the recursion method of finite systems up to 24 sites.
Two types of the exchange interaction are considered: one is the nearest-neighbor type, and the other is the inverse-square one.
As the Ising anisotropy becomes larger, there appears a noticeable difference in the transverse component $S^{xx}(q,\omega)$ between the two types of the exchange.
For the nearest-neighbor type, the peak frequency of $S^{xx}(q,\omega)$ for each $q$ approaches the center of the continuum spectrum.
On the contrary, the peak frequency for the inverse-square type moves to the upper edge of the continuum, and separates from the continuum for the anisotropy larger than the threshold value.
Whether the interaction between domain walls (solitons) is absent or repulsive in the Ising limit leads to this difference in the behavior of $S^{xx}(q,\omega)$.
In the longitudinal component $S^{zz}(q,\omega)$, on the other hand, the feature of the dynamics is scarcely different between the two types.
The energy gap and the static properties are also discussed.
}

\kword
{
dynamical structure factor, $XXZ$ chain, inverse-square exchange, nearest-neighbor exchange, Ising anisotropy, Haldane-Shastry model, exact diagonalization, recursion method
}

\begin{document}
\sloppy
\maketitle



\section{Introduction}

The physics concerning one-dimensional (1D) quantum systems has been a long-standing problem.
Discovery of many quasi-1D quantum magnets has spurred on the studies of the systems.
Among those, there are $S=1/2$ systems in which a finite energy gap opens in the magnetic excitation spectrum.
These systems have various origins of an excitation gap.
For example, in CuGeO$_3$, the first inorganic compound in which a spin-Peierls transition was observed,~\cite{HTU} a gap appears below the critical temperature because of the lattice dimerization and the frustration.
The other case with a gap is the $XXZ$ chain with Ising anisotropy.
CsCoCl$_3$ is known to be a typical 1D Ising-like antiferromagnet.~\cite{Achiwa}One of the experimental means to measure the spin gap is the inelastic neutron scattering.~\cite{NFA,HY}
The neutron scattering brings us not only the information of the spin gap but also the whole excitation spectrum.~\cite{AFMAB,YHSS}
In this context, the dynamical structure factor (DSF) is a physical quantity to investigate carefully, because it is closely related with the experimental data detected by the neutron scattering.
\par

Besides connection with experiment, the study of dynamics is indispensable in order to understand the nature of elementary excitations in the system.
To this end, it is desirable to find the exact expression of the DSF.
Haldane and Zirnbauer~\cite{HZ} achieved this great work in the Haldane-Shastry (HS) model,~\cite{Haldane,Shastry} which is an isotropic spin model with exchange interaction proportional to the inverse square of the distance.
Their result has the biggest merit that the picture of the elementary excitation is made simple.
Recently, an integral representation of the exact DSF has been clarified for the nearest-neighbor Heisenberg model,~\cite{BCK,Bougourzi} although the result is considerably intricate.
It is known that the HS model and the Heisenberg model belong to the same universality class, and that the DSFs for these models are similar to each other.
\par

How do the features of the dynamics change when the Ising anisotropy is introduced in these models?
This problem on the $XXZ$ model is linked with formation of the finite gap.
For the nearest-neighbor exchange, there exist a perturbation theory from the Ising limit~\cite{IS} and calculation based on the spin wave continuum.~\cite{MM}
To our knowledge, however, there is no work to study how the dynamics changes as the Ising anisotropy is increased.
Moreover, for the inverse-square $XXZ$ model, systematic studies of the dynamical properties have been lacking.
\par

The purpose of this paper is to clarify the effects of the Ising anisotropy and the interaction range on the DSFs for the 1D $XXZ$ model. 
Two types of the exchange are considered: the nearest-neighbor exchange and the inverse-square one.
We calculate the DSFs up to 24 sites via the exact diagonalization and the recursion method.~\cite{GB}
We elucidate the change of the dynamics due to the Ising anisotropy and the difference between the two types of the exchange.
Consequently, it is found that for the system with the large Ising anisotropy, the dynamics is sensitive to the range of the exchange interaction.
In addition, the ground-state properties, the energy gap and the static structure factor (SSF) are investigated.
The energy gap and the SSF can be extracted from the process of calculation for the DSF.
As a result, the characteristic difference is found in the opening of the energy gap between the two types of the exchange.
\par

The remainder of this paper is organized as follows. 
In the next section ($\S\ref{Model}$) we define the Hamiltonian of the 1D $XXZ$ model and the state vector.
We review the recursion method in $\S\ref{Method}$.
Then, we show the results for the inverse-square $XXZ$ model, comparing with those for the nearest-neighbor model in $\S\ref{LR}$.
In $\S\ref{Interpretation}$, we discuss the behavior of the DSF from the Ising limit, and formulate the energy levels for the inverse-square model by using the two-soliton approximation.
Finally we give concluding remarks in $\S\ref{Conclusion}$. 
\par


\section{Model} \label{Model}

We consider the $S=1/2$ 1D antiferromagnetic $XXZ$ model
\be
  {\cal H} = \sum_{i<j}^N J_{ij} \bigl( S_i^x S_j^x + S_i^y S_j^y + \Delta S_i^z S_j^z \bigr),\label{Hamiltonian}
\ee
where $N$ is the number of spins which is assumed to be even, and $J_{ij}$ denotes the exchange interaction between spins $i$ and $j$.
The parameter $\Delta$ $(\ge 1)$ represents the Ising anisotropy.
The periodic boundary condition is imposed on the system.
In this paper, we take two types of the exchange interaction:

\noindent(i) the nearest-neighbor (NN) type, i.e., $J_{ij}=J\delta_{j,i+1}$;

\noindent(ii) the inverse-square (IS) type, i.e., $J_{ij}=JD(x_i-x_j)^{-2}$, where $D(x_i-x_j)=(N/\pi)\sin[\pi(x_i-x_j)/N]$.

\noindent The Hamiltonians in the cases (i) and (ii) are written as ${\cal H}_{\rm NN}$ and ${\cal H}_{\rm IS}$, respectively.
\par

States in the Hilbert space for a spin system can be represented by the position of up and down spins.
If $M$ denotes the number of down spins, the $z$-component of the total spin is given by
\be
  S_{\rm tot}^z = \sum_{i=1}^N S_i^z = \frac{N}{2}-M.
\ee
This quantity can be a good quantum number of the system.
A state vector in the subspace with $S_{\rm tot}^z$ fixed is represented by
\be
  |\psi \rangle = \sum_{\{x\}} \psi(\{x\}) \prod_{i=1}^M S_{x_i}^- |F \rangle,\label{statevector}
\ee
where $|F \rangle$ denotes the fully polarized up-spin state, and $S_{x_i}^-$ is the spin-lowering operator at site $x_i$.
The wave function $\psi(\{x\})$ is symmetric in the position $\{x\}=\{x_1,\cdots,x_M\}$ of the down spins.
In eq.\ (\ref{statevector}), the number of the bases is ${}_N C_M$.
Because the ground state of the system belongs to the subspace with $S_{\rm tot}^z=0$ for $\Delta \ge 1$, we can take the lowest state obtained by diagonalizing within the subspace $S_{\rm tot}^z=0$.
The information of the ground state is necessary to calculate the DSF at zero temperature.
\par


\section{Numerical Method} \label{Method}

The DSF $S^{\mu\mu}(q,\omega)$ is described as the Fourier transform of the dynamical correlation function $\langle 0| S_{\ell}^{\mu}(t) S_{\ell+R}^{\mu} |0 \rangle$, i.e.,
\bea
  S^{\mu\mu}(q,\omega) &=& \frac{1}{N} \sum_{\ell=1}^N \sum_{R=0}^{N-1} {\rm e}^{-{\rm i}qR} \int_{-\infty}^{\infty} \frac{dt}{2\pi} {\rm e}^{{\rm i} \omega t} \langle 0 |S_{\ell}^{\mu}(t) S_{\ell+R}^{\mu}| 0 \rangle \nonumber \\
                       &=& \sum_n M^{\mu\mu}(q,\omega_n) \,\delta \left( \omega - \omega_n \right),\label{Sqomega2} \\
  M^{\mu\mu}(q,\omega_n) &=& \left| \langle n | S_q^\mu | 0 \rangle \right|^2,\label{Mqomega} \\
  \omega_n &=& E_n - E_0,
\eea
where $\mu=x,y$ or $z$, $S_q^{\mu} = N^{-1/2} \sum_\ell {\rm e}^{-{\rm i}q \ell} S_\ell^{\mu}$ ($q$ is a momentum transfer), 
and $|n\rangle$ denotes an eigenstate of ${\cal H}$ with energy $E_n$ ($E_0$ being the ground-state energy).
$S^{\mu\mu}(q,\omega)$ is extracted from the Green function $G^{\mu\mu}(q,z)$ as follows:~\cite{GB}
\bea
  S^{\mu\mu}(q,\omega) &=& - \frac{1}{\pi} \lim_{\eta \to 0} {\rm Im} G^{\mu\mu}(q,\omega+i\eta+E_0),\label{Sqomega3} \\
  G^{\mu\mu}(q,z) &=& \langle 0| S_{-q}^{\mu} (z-{\cal H})^{-1} S_q^{\mu} |0 \rangle.\label{Green}
\eea
The Green function $G^{\mu\mu}(q,z)$ is written in the form of a continued fraction:
\be
  G^{\mu\mu}(q,z) = \cfrac{S^{\mu\mu}(q)}{z-a_0-
                    \cfrac{b_1^2}{z-a_1-
                    \cfrac{b_2^2}{z- \cdots}}},\label{Green2}
\ee
where $S^{\mu\mu}(q) \equiv \langle 0|S_{-q}^\mu S_q^{\mu}|0 \rangle$ means the SSF.
The coefficients $a_n$ and $b_n$ are obtained from the Lanczos algorithm in which we set the normalized initial vector as $|f_0 \rangle = S_q^{\mu}|0 \rangle / || S_q^{\mu}|0 \rangle ||$,
\bea
  |f_{n+1} \rangle &=& b_{n+1}^{-1} [({\cal H}-a_n)|f_n \rangle - b_n|f_{n-1} \rangle],\nonumber \\
  a_n &=& \langle f_n|{\cal H}|f_n \rangle,\nonumber \\
  b_{n+1} &=& ||({\cal H}-a_n)|f_n \rangle - b_n|f_{n-1} \rangle ||, \quad b_0=0.\label{relation}
\eea
Note that $|f_{n+1} \rangle$ is normalized by $b_{n+1}$.
\par

The nonzero components of the DSF are $S^{xx}(q,\omega)$, $S^{yy}(q,\omega)$ and $S^{zz}(q,\omega)$.
Because the Hamiltonian eq.\ (\ref{Hamiltonian}) has the axial symmetry, the transverse components $S^{xx}(q,\omega)$ and $S^{yy}(q,\omega)$ are related with $S^{+-}(q,\omega)$ and $S^{-+}(q,\omega)$ as follows:
\be
  S^{xx}(q,\omega) = S^{yy}(q,\omega) = \frac{1}{2} S^{+-}(q,\omega) = \frac{1}{2} S^{-+}(q,\omega).
\ee
In substance, we have only to calculate two quantities: the longitudinal component $S^{zz}(q,\omega)$ and the transverse one $S^{+-}(q,\omega)$ $[=2S^{xx}(q,\omega)]$.
\par

Thus the procedure for calculation of the DSF is summarized as follows:

(1) We calculate the ground-state energy $E_0$ and eigenvector $|0\rangle$ via the Lanczos method and the inverse iteration one.
In this case, we can take an arbitrary vector as the initial vector, except a vector orthogonal to the ground-state eigenvector.

(2) We make $|f_0 \rangle = S_q^{\nu}|0 \rangle / ||S_q^{\nu}|0 \rangle||$ ($\nu = z$ or $-$), and evaluate the coefficients $a_n$ and $b_n$ by using eq.\ (\ref{relation}).

(3) Once we obtain $a_n$ and $b_n$, we can construct $S^{zz}(q,\omega)$ or $S^{xx}(q,\omega)$ from eqs.\ (\ref{Sqomega3}) and (\ref{Green2}).
\par

Originally the number of the coefficients $a_n$ and $b_n$ should be equal to the one of poles at each momentum $q$.
For our systems, however, the results in the low-energy region are not sensitive to the number of coefficients used for $n \ge 30$.
We calculate with use of at least 30 coefficients $a_n$ and $b_n$.
\par

In the practical numerical calculation, one needs to take $\eta$ as a small but finite value.
This means that the $\delta$-function is approximated by the Lorentzian.
Thus $\eta$ corresponds to the Lorentzian width.
For the purpose of inspecting the distribution of poles in detail, one should choose the value of $\eta$ as close as possible to zero.
We use $\eta = O (10^{-5} J)$ on taking account of computing time.
\par

Finally, we check the numerical precision by using the sum rule
\be
  \int_{-\infty}^\infty d\omega \frac{S^{\mu\mu}(q,\omega)}{S^{\mu\mu}(q)} = 1.\label{sumrule}
\ee
All the data in this paper reach the precision more than 0.999 for every momentum $q$.
However, we note that the values of poles in the high-energy region are not reliable in contrast to the excellent accuracy with the low-lying states, as the previous papers~\cite{GB,HRD,Takahashi} pointed out. 
\par

In this way, we can obtain $\omega_n$ and $M^{\mu\mu}(q,\omega_n)$ for each $q$ in finite size $N$.
In order to extract the behavior of $S^{\mu\mu}(q,\omega)$ in the thermodynamic limit from finite-size data, we use the Lanczos spectra decoding (LSD) method.~\cite{ZS}
The procedure gives the DSF as follows:
\bea
  S^{\mu\mu}(q,\bar{\omega}_n) &=& \frac{1}{2} \cdot \frac{M^{\mu\mu}(q,\omega_{n+1}) + M^{\mu\mu}(q,\omega_n)}{\omega_{n+1} - \omega_n},\quad \\
  \bar{\omega}_n &=& \frac{\omega_n + \omega_{n+1}}{2}.
\eea
However, this method is not easy to apply to cases where there exists an isolated mode with $\delta$-function contribution, and/or different kinds of continua overlap (for instance, a two-spinon continuum and a four-spinon one).
If it is possible to separate each contribution from the raw data, we can use the remaining data to apply the LSD method.
\par


\section{Results for the Inverse-Square Exchange and Comparison with the Nearest-Neighbor Exchange} \label{LR}

We focus on the following physical quantities: the ground-state energy, the energy gap, the DSF and the SSF.
For the IS exchange, the completely isotropic problem ($\Delta=1$) has been solved analytically.
After recalling the analytic expressions in each subsection, we present our results numerically obtained for $\Delta>1$ and discuss how these deviate from the ones for $\Delta=1$. 
For comparison, we show the results for the NN $XXZ$ model.
Then we can bring out the features of the IS $XXZ$ model in full relief.
\par


\subsection{Ground-state properties}

In the IS isotropic case, the following simple form (i.e., the Jastrow-type function) yields the exact ground-state wave function:~\cite{Haldane,Shastry}
\be
  \psi_{\rm J}(\{x\}) = \prod_{i=1}^{N/2} (-1)^{x_i} \prod_{i<j}^{N/2} |D(x_i - x_j)|^2.
\ee
Its eigenenergy $E_0$ is derived as follows:
\bea
  \frac{E_0}{NJ} &=& - \frac{\pi^2}{24} \left( 1+\frac{5}{N^2} \right) \label{gseisotropicis} \\
                 &\stackrel{N \to \infty}{\longrightarrow}& - \frac{\pi^2}{24} = -0.41123 \cdots .\label{HSexactsolution}
\eea
\par

For the NN isotropic model, the ground-state energy is known as~\cite{Bethe,Hulthen}
\be
  \frac{E_0}{NJ} = -\ln 2 + \frac{1}{4} = -0.44314 \cdots .\label{gseisotropic}
\ee
This value is a little smaller than that for the IS isotropic model given by eq.\ (\ref{HSexactsolution}).
\par

In the following we investigate the anisotropic case. 
First, let us look at the ground-state energy for the IS model with $\Delta > 1$. 
We extrapolate the values in the thermodynamic limit by using the data from $N=$6 to 24 and the following polynomial:
\be
  \Gamma(N) = \Gamma(\infty) + c_1 \frac{1}{N} + c_2 \frac{1}{N^2},\label{polynomial}
\ee
where $\Gamma = E_0 / (N J)$.
In Fig.\ 1, we show the values evaluated for $\Delta = 2, 5$ and 10 and interpolate those.

For the NN model, Walker expanded the result of eq.\ (\ref{gseisotropic}) into the anisotropic case ($\Delta>1$).~\cite{Walker}
His result is equivalent to the following expression, which was later obtained by des Cloizeaux and Gaudin:~\cite{dCG}
\be
  \frac{E_0(\Delta)}{NJ} = -\sinh \Psi \left[ \sum_{n=1}^\infty (1-\tanh n \Psi) + \frac{1}{2} \right] + \frac{1}{4} \cosh \Psi,\label{gseanisotropic}
\ee
where $\Delta = \cosh \Psi$ and $0<\Psi<+\infty$. 
The limit $\Delta \to 1$ (i.e., $\Psi \to 0$) of eq.\ (\ref{gseanisotropic}) coincides with eq.\ (\ref{gseisotropic}).
In Fig.\ 1, we show the $\Delta$-dependence of the ground-state energy by the thin solid line.
This figure indicates that the difference of the ground-state energy between the IS exchange and the NN one widens as the Ising anisotropy becomes larger.
\par

Next, we consider the wave function.
Because eqs.\ (\ref{gseisotropic}) and (\ref{gseanisotropic}) for the NN model are derived on the basis of the Bethe ansatz, the corresponding wave functions are very complex.
From here, we concentrate the wave function for the IS $XXZ$ model.

In this case, the Jastrow-type function
\be
  \psi_{\rm J}(\{x\}) = \prod_{i=1}^M (-1)^{x_i} \prod_{i<j}^{M} |D(x_i - x_j)|^\lambda,\label{wf}
\ee
is the exact ground-state wave function in the subspace of the number $M$ of down spins under the conditions
\be
  \lambda (M-1) \le N,
\ee
and $\lambda= $even integer, where $\Delta = \lambda ( \lambda - 1 )/2$.~\cite{Haldane}
When $\Delta > 1$ (i.e., $\lambda > 2$), this wave function eq.\ (\ref{wf}) is {\it not} the absolute ground state (i.e., $M=N/2$) of the system. 
In order to test the accuracy of this wave function as a trial function, the overlap $|\langle \psi_{\rm exact} | \psi_{\rm J} \rangle|$ between eq.\ (\ref{wf}) and the exact eigenfunction $\psi_{\rm exact}$ is calculated by the numerical diagonalization. 
\par

In our previous paper,~\cite{SKK} we showed that the overlap is larger than 0.95993 (of course, smaller than unity) over whole range of the Ising anisotropy ($\Delta > 1$) for $N=12$ and $M=6$.
We extend this study up to $N=22$. 
Figure 2 shows the overlap against the Ising anisotropy for 22 sites.
In this case, the overlap is not as close to unity as that for 12 sites, 
but is still more than 0.88305.
The $\Delta$-dependence of the overlap has a conspicuous minimum at a certain value of $\Delta$.
This value of $\Delta$ sweeps toward the smaller one as the size $N$ becomes larger.
For $N=6, 12$ and 22, the integer values of $\Delta$ with the minimum overlap are $\Delta=$8, 6 and 5, respectively.
\par

One may naturally ask the size dependence of the overlap for fixed Ising anisotropy. 
We show the cases $\Delta = 2, 5$ and 10 in Fig.\ 3.
In all cases, the overlaps decay with increasing size $N$.
However, the decaying rate is different: it is most rapid for $\Delta = 5$.
For $\Delta = 1$ and $\Delta \to \infty$, the Jastrow-type function eq.\ (\ref{wf}) represents the exact eigenstate, as mentioned in ref.\ 22.
The results in Fig.\ 3 suggest that the overlaps for $1 < \Delta < \infty$ vanish in the limit $N \to \infty$. 
Nevertheless the wave function eq.\ (\ref{wf}) is surely an excellent trial function for the finite systems, at least with $N \le 22$. 
\par


\subsection{Energy gap}

To compare with new results for the IS model, first we review the result for the NN model.
Des Cloizeaux and Gaudin investigated the nature of the ground state and the first excited states for all values of $\Delta$.~\cite{dCG}
Particularly, they demonstrated that for $\Delta>1$, an energy gap $E_{\rm G}(\Delta)$ appears, which can be defined as,
\be
  E_{\rm G}(\Delta) = \lim_{N \to \infty} \left[ E(\Delta,1,q) - E(\Delta,0,q) \right].
\ee
Here $E(\Delta, S_{\rm tot}^z, q)$ means the lowest energy in the subspace with two quantities $(S_{\rm tot}^z, q)$ fixed for the anisotropy parameter $\Delta$.
The expression of the energy gap is given by
\be
  E_{\rm G}(\Delta) = \frac{\pi J \sinh \Psi}{\Psi} \sum_{n=-\infty}^{+\infty} \frac{1}{\cosh [(2n+1)\pi^2/(2\Psi)]},\label{NNgap}
\ee
where $\Delta = \cosh \Psi$. 
This gap increases very slowly at the threshold as follows:
\bea
  E_{\rm G}(\Delta) &\simeq& 4 \pi J \exp \left[ - \frac{\pi^2/(2 \sqrt{2})}{\sqrt{\Delta - 1}} \right],\nonumber \\
                    & & 0 < \Delta - 1 \ll 1.
\eea
Therefore the energy gap has the essential singularity at the critical point $\Delta = 1$; all differentials become zero at the point.
When $\Delta$ goes to infinity, the gap behaves as
\be
  E_{\rm G}(\Delta) \simeq J (\Delta - 2).\label{asymptote}
\ee
In Fig.\ 4, we show $E_{\rm G}$ as a function of $\Delta$ by the thin solid line.
\par

Next, we numerically investigate the behavior of the gap for the IS $XXZ$ model.
The excitation gap is defined as,
\be
  E_{\rm G}^{q^\ast}(\Delta) = \lim_{N \to \infty} [E(\Delta, 1, q^\ast) - E(\Delta, 0, 0)].
\ee
We consider the cases $q^\ast = 0$ and $\pi$.
\par

In Figs.\ 5(a) and 5(b), we show the results of finite-size analysis with the data of $N=6$-24 and the polynomial eq.\ (\ref{polynomial}) [$\Gamma = E(\Delta, 1, q^\ast) - E(\Delta, 0, 0)$].
In some cases the extrapolated value is slightly negative.
For instance, the value of $E_{\rm G}^0$ is $-0.030023$ for $\Delta = 2$.
These cases with a negative gap indicate the degree of numerical inaccuracy.
The $\Delta$-dependence of the gap is shown in Fig.\ 4.
\par

We find the following features from Fig.\ 4.
First, for the IS type, the opening of the gap is slower than that for the NN one.
Second, both $E_{\rm G}^0$ and $E_{\rm G}^\pi$ for the IS model seem to be well fitted by the essentially singular form.~\cite{Hatsugai}
However, it is difficult to evaluate the quantitative form by our numerical calculation: we cannot determine the accurate critical point.
\par


\subsection{Excitation spectra and dynamical structure factors} \label{ISESDSF}

In the IS isotropic case (i.e., the HS model), Haldane and Zirnbauer~\cite{HZ} derived the exact expression of the DSF in the thermodynamic limit. 
It is given by
\bea
  &&S^{xx}(q,\omega) = S^{zz}(q,\omega) \nonumber \\
  &&= \frac{1}{4} \frac{ \Theta( \omega_{\rm U}(q) - \omega ) \Theta( \omega - \omega_{{\rm L}-}(q) ) \Theta( \omega - \omega_{{\rm L}+}(q) )}{[ (\omega - \omega_{{\rm L}-}(q)) (\omega - \omega_{{\rm L}+}(q)) ]^{1/2}},\qquad \label{exactSqomega}
\eea
where $\Theta(\omega)$ is the step function, and
\bea
  \omega_{{\rm L}-}(q) &=& \frac{J}{2} q (\pi - q), \label{omegaL-} \\
  \omega_{{\rm L}+}(q) &=& \frac{J}{2} (q - \pi) (2\pi - q), \label{omegaL+} \\
  \omega_{\rm U}(q) &=& \frac{J}{4} q (2\pi - q). \label{omegaU}
\eea
Equations (\ref{omegaL-}) and (\ref{omegaL+}) represent the lower edge of the excitation spectrum for $0 \le q \le \pi$ and for $\pi \le q \le 2\pi$ respectively, and eq.\ (\ref{omegaU}) gives its upper boundary for $0 \le q \le 2\pi$.
The intensity diverges at the lower edge of the continuum spectrum and jumps to zero at the upper edge. 
\par

Here we perform the numerical calculation of the DSF in the HS model.
The same calculation has been done up to 16 sites.~\cite{TH}
We show our result for $N=24$ in Fig.\ 6.
It reveals clearly the compact support for the excitation; there is no intensity outside of the continuum.~\cite{FLS}
This is due to the selection rule of the Yangian, which is a special symmetry with the HS model.
Namely the DSF consists only of the two-spinon excitations.~\cite{HZ,TH}
\par

In this context, the situation is more complicated for the NN isotropic model (i.e., the Heisenberg model).
For this model, the higher-order-spinon processes as well as the two-spinon ones contribute to the DSF.~\cite{Bougourzi,MTBB,KMBFM}
The exact two-spinon DSF has been obtained by an integral representation.~\cite{BCK}
\par


We test the LSD method~\cite{ZS} in the HS model.
In Fig.\ 7, we compare the numerical result (closed symbols) processed by the LSD method for $N=$16-24 with the exact one (solid line) of eq.\ (\ref{exactSqomega}) at $q=\pi$, which becomes
\bea
  &&S^{xx}(\pi,\omega) = S^{zz}(\pi,\omega) \nonumber \\
  &&= \left\{
   \begin{array}{ll}
    1 / (4 \omega) \quad &\mbox{for $0 \le \omega/J \le \pi^2/4$},\\
    0 \quad &\mbox{otherwise}.
   \end{array}\right.\label{HSsqwpi}
\eea
It is found from Fig.\ 7 that the numerical result agrees fairly well with the exact one, especially near the upper boundary.
There appears clearly the step-function-like behavior peculiar to the HS model.
The numerical result deviates somewhat from the exact one in the lower-energy region where the number of poles is fewer (that is, the density of states is smaller).
The deviation occurs because the finite difference is substituted for the differential in the LSD method.~\cite{ZS}
\par

Now, let us turn to the anisotropic case.
We show $S^{xx}(q,\omega)$ for the IS $XXZ$ model with $\Delta = 2, 5$, and 10 in Figs.\ 8(a), 8(b) and 8(c), respectively.
Characteristic features with increasing Ising anisotropy are:

[A] The shape of the continuum tends to be symmetric about $q/\pi=1/2$ but the intensity does not.

[B] The main intensity shifts from the low-energy region to the high-energy one of the continuum.
\par

Encouraged by the validity of the LSD method for the HS model, we apply the method to anisotropic cases.
Unlike for $\Delta = 1$, first we select poles with main intensity ($> 10^{-2}$), which are presumably due to two-soliton contribution, and subsequently process the selected data by the LSD method.
Figure 7 shows the results at $q=\pi$ for $N =$16-24 and $\Delta=1, 2$ and 5.
The data for $(N,\Delta) = (18,2)$ have been removed since some numerical inaccuracy causes the irregularity of the dependence on $N$.
This figure makes the above feature [B]  clear.
Note that the peak intensity does not shift gradually inside the continuum; it appears either at the lower boundary or at the upper one.
\par

For comparison, we apply the same method to the NN model.
At $q=\pi$, the number of poles with main intensity we have selected is equal to the number of all poles for the HS model in each $N$.
In Fig.\ 9 we show the change of $S^{xx}(\pi,\omega)$ as a function of $\Delta$.
For $\Delta = 1$, $S^{xx}(\pi,\omega)$ has a rounded shoulder at the upper boundary~\cite{KMBFM} in contrast to the HS model.
Moreover, unlike the IS case, $S^{xx}(\pi,\omega)$ which forms the continuum becomes roundish with increasing $\Delta$.
The peak frequency approaches the center of the continuum.
\par

What happens in the behavior of the DSF  when the Ising anisotropy is much larger in both the IS $XXZ$ model and the NN one?
We show the result for the IS model with $\Delta=10$ in Fig.\ 10.
To see the size dependence of poles and intensities, we show the behavior of $M^{xx}(\pi,\omega)$ [see eq.\ (\ref{Mqomega})] in the main plot of Fig.\ 10(a).
The following features are observed:
\begin{itemize}
\item Among poles with main intensity (which all are indicated in this figure for $\Delta=10$), the highest pole position is hardly dependent on size $N$, in contrast with other poles.
\item The intensity of the highest pole oscillates with $N$ but seems to converge to a certain value.
\end{itemize}
From these features, it is likely that the highest pole is an isolated mode separated from the continuum to which the other poles belong.
For the purpose of judging the separation, it is useful to see the behavior of $NM^{xx}(\pi,\omega)$, shown in the inset of Fig.\ 10(a).
This figure strengthens the possibility of the isolated pole.
We regard the highest pole as the $\delta$-function contribution.
Then we eliminate this pole and analyze remaining poles by the LSD method, which results in the continuum part of $S^{xx}(\pi,\omega)$ shown in Fig.\ 10(b).
It seems to be a monotonically increasing function of $\omega$.
On the other hand, there appears no isolated mode for the NN model with $\Delta=10$, which is consistent with the result by the perturbation approach from the Ising limit.~\cite{IS}
\par

Because there exists the isolated mode for the IS model with $\Delta=10$, one may naturally ask the threshold value $\Delta_{\rm c}$ at which the mode begins to appear.
We judge that the isolated mode exists if the behavior of $M^{xx}(\pi,\omega)$ [or $NM^{xx}(\pi,\omega)$] satisfies the following feature in addition to the two itemized above:
two-soliton contribution with main intensity does not overlap in higher-order-soliton one with very small intensity.
See Figs.\ 11(a) and 11(b), where we show the results for $\Delta=6$ and 7 respectively.
We infer $6 < \Delta_{\rm c} \, \lsim \, 7$ from these figures.
In order to determine the precise value of $\Delta_{\rm c}$, it is necessary to perform detailed analyses of poles and intensities in systems with still larger size.
\par

Next, we show $S^{zz}(q,\omega)$ for the IS model with $\Delta = 2$ and $5$ in Figs.\ 12(a) and 12(b), respectively.
For $S^{zz}(q,\omega)$, the character of the dynamics is scarcely different from that for the NN model.
Instead of [B], the peculiarity to $S^{zz}(q,\omega)$ comes out:
[B'] The intensity at $\omega \simeq 0$ and $q=\pi$ is conspicuously enhanced as the Ising anisotropy becomes larger.
It is because the N\'eel-type correlation strengthens at $q=\pi$.
Meanwhile, the other intensities are suppressed.
For $\Delta=5$, the poles with weak intensity seem to belong to a continuum.
This continuum may have the feature [A] on the analogy of the result obtained by Ishimura and Shiba.~\cite{IS}
\par


\subsection{Static structure factors}

Before the IS isotropic model (i.e., the HS model) was proposed, Gebhard and Vollhardt~\cite{GV} calculated the expression of the SSF in terms of the Gutzwiller wave function~\cite{Gutzwiller} as follows:
\be
  S^{xx}(q) = S^{zz}(q) = - \frac{1}{4} \ln \left( 1 - \frac{q}{\pi} \right),\label{sqisotropicl}
\ee
for $0 \le q \le \pi$.
After that, it was demonstrated that the ground-state wave function of the HS model is identical to the Gutzwiller one.~\cite{Haldane,Shastry}
Thus eq.\ (\ref{sqisotropicl}) turns out to be the exact expression of the SSF for the HS model.
This was confirmed by carrying out the integration of eq.\ (\ref{exactSqomega}) over $\omega$.~\cite{HZ}
From eq.\ (\ref{sqisotropicl}), the SSF is proportional to $q$ for small $q$, and it is logarithmically divergent at $q=\pi$.
\par

In Fig.\ 13, we compare this exact result (solid line) with our numerical data (closed symbols) for the HS model ($\Delta = 1$).
The fit of the symbols to eq.\ (\ref{sqisotropicl}) is very good except at $q=\pi$.
The data at $q=\pi$ are noticeably dependent on the size $N$, reflecting the logarithmic divergence.
\par

Let us turn now to the anisotropic case.
First, we look at the transverse component $S^{xx}(q)$.
Figure 13 shows $S^{xx}(q)$ for the IS model with $\Delta = 2$ and 5 besides $\Delta = 1$.
The $q$-dependence of $S^{xx}(q)$ becomes weaker with increasing Ising anisotropy.
$S^{xx}(q)$ is independent of $q$ in the Ising limit; the value of $S^{xx}(q)$ is equal to 1/4.
\par

Next, we show the longitudinal component $S^{zz}(q)$ for the IS model with $\Delta = 1, 2$ and 5 in Fig.\ 14.
As the Ising anisotropy becomes larger, $S^{zz}(q)$ at $q=\pi$ rapidly increases; $S^{zz}(\pi)$ has considerable dependence on size $N$.
On the other hand, $S^{zz}(q)$ for $q < \pi$ are suppressed.
This leads to the weak intensity in $S^{zz}(q,\omega)$ for $0 < q <\pi$.
\par


\section{Interpretation of $S^{xx}(q,\omega)$ from the Ising Limit} \label{Interpretation}

\subsection{Qualitative interpretation} \label{Qualitative}

What is the origin of different behavior of $S^{xx}(q,\omega)$ for $\Delta \gg 1$ between the IS $XXZ$ model and the NN one?
We consider this question from the Ising limit for both models. 
\par

First, let us take notice of the energy levels. 
In the Ising limit, the ground-state level is doubly degenerate corresponding to the N\'eel states with a different choice of sublattice for both models (see ref.\ 30 and Appendix). 
However, the nature of the next level differs between the two models. 
Here we limit the Hilbert space to the subspace with $S_{\rm tot}^z = \pm 1$, in order to inspect the levels related with $S^{xx}(q,\omega)$. 
For the NN case, the next level is separated from N\'eel states by $J$ and is highly degenerate.
The states with this level have only two domain walls, or solitons.
We can interpret as free solitons.
On the other hand, for the IS type, the degeneracy is removed except that due to the translational symmetry.
We find numerically that the first excited state is the state which has two solitons with the longest distance; moreover, among the excited states with two solitons, the state with the nearest-neighbor solitons has the highest energy (see Appendix).
This means that {\it two solitons are repulsive for the IS Ising model}. 
\par

Next, we consider the transverse component $S^{xx}(q,\omega)$ of the DSF.
See the results of Appendix.
In both models, the states contributing to $S^{xx}(q,\omega)$ are limited to those with the nearest-neighbor solitons.
As mentioned above, however, the energy level of this state is equal to that of other states with two solitons for the NN Ising model, whereas it is the highest among the energy levels of the two-soliton states for the IS model.
\par

When the spin-flip terms are introduced in the Hamiltonian, the energy level is split up and down for the NN model.
On the contrary the levels for the IS model are only modified a little because these are not degenerate from the beginning.
The DSFs come to contribute to many states in both models.
Nevertheless each nature in the Ising limit remains if the spin-flip terms are much smaller than the Ising term: most of the intensity comes from the vicinity of the energy level with the nearest-neighbor-soliton state in the Ising limit.
\par

In this way, the behavior of $S^{xx}(q,\omega)$ for $\Delta \gg 1$ is qualitatively explained.
\par

\subsection{Two-soliton problem in the inverse-square $XXZ$ model}

In order to treat the discussion on the energy levels quantitatively, we study the excitation spectrum for the IS $XXZ$ model by using the two-soliton approximation.
This approximation should be valid for description of the low-lying excitation near the Ising limit.
\par

We redefine the Hamiltonian as follows:
\be
  {\cal H}_{\rm IS}/\Delta \equiv \tilde{{\cal H}}_{\rm IS} = \sum_{i<j}^N J_{ij} \left[ \frac{\epsilon}{2} \left( S_i^+ S_j^- + S_i^- S_j^+ \right) + S_i^z S_j^z \right],\label{HamiltonianISD}
\ee
where $J_{ij} = J D(x_i - x_j)^{-2}$.
The two-soliton eigenstate is written by a linear combination of all possible configurations of two domain walls from the reference state $|{\rm Neel}_1 \rangle \equiv |\uparrow \downarrow \uparrow \downarrow \cdots \uparrow \downarrow \rangle$, i.e.,
\bea
  |\psi \rangle &=& \sum_i f_{ii} S_i^+ |{\rm Neel}_1 \rangle \nonumber \\
                &+& \sum_{i \ne j} \left[ \Theta(x_j - x_i) f_{ij} S_i^+ S_{i+1}^- \cdots S_j^+ |{\rm Neel}_1 \rangle \right. \nonumber \\
                &+& \left. \Theta(x_i - x_j) f_{ij} S_i^+ S_{i+1}^- \cdots S_N^+ S_1^- \cdots S_j^+ |{\rm Neel}_1 \rangle \right], \nonumber \\
\eea
where $\Theta(x)$ is the step function.
Note that both subscripts $i$ and $j$ of $f_{ij}$ run over only even numbers.
Thus the total degree of freedom becomes $(N/2)^2$.
\par

Let us consider the eigenvalue problem 
\be
  \tilde{{\cal H}}_{\rm IS} |\psi \rangle = E |\psi \rangle. \label{ep}
\ee
First, we define a relative coordinate $\bar{r}$ between the two solitons as follows:
\be
  \bar{r} = \left\{
   \begin{array}{ll}
    x_j - x_i \quad &\mbox{for $x_j \ge x_i$},\\
    \left( x_j + N \right)- x_i \quad &\mbox{for $x_j < x_i$}.
   \end{array}\right.\label{relativecoordinate}
\ee
Then the permissible values of $\bar{r}$ are limited to $0, 2, 4, \cdots, N-2$.
By taking the inner product of both sides of eq.\ (\ref{ep}) with every basis state $\langle {\rm Neel}_1 | S_j^- \cdots S_{i+1}^+ S_i^-$, we obtain the following equation for the amplitudes $f_{ij}$:
\bea
  && \left[ E - E_0 - E_{{\rm add}(\bar{r}+1)} \right] f_{ij} \nonumber \\
  && - \frac{\epsilon}{2} J(1) \left[ \left( 1 - \delta_{j,i} \right) \left( f_{i+2,j} + f_{i,j-2} \right) \right. \nonumber \\
  && \left. + \left( 1 - \delta_{j,i-2} - \delta_{j,i+N-2} \right) \left( f_{i,j+2} + f_{i-2,j} \right) \right] = 0,\label{equationfij}
\eea
where
\bea
  && E_0 = - \frac{\pi^2}{48} J N \left( 1 + \frac{2}{N^2} \right),\label{ISIgsenergy} \\
  && E_{{\rm add}(\bar{r}+1)} = - \frac{1}{2} J \left( \frac{\pi}{N} \right)^2 \nonumber \\
  && \hspace{1cm} \times \sum_{i=1}^{\bar{r}+1} \sum_{j=\bar{r}+2}^{N} \frac{(-1)^{x_i-x_j}}{\sin^2 [\pi(x_i-x_j)/N]},\\
  && J(1) = J \left( \frac{\pi}{N} \right)^2 \frac{1}{\sin^2 (\pi/N)}.
\eea
\par

Next, we separate out the center of mass motion.
Beside the definition of the relative coordinate $\bar{r}$ by eq.\ (\ref{relativecoordinate}), we introduce the center of gravity $\bar{R}$, the total momentum $q$ and relative momentum $k$,
\bea
  && 2 \bar{R} = \left\{
   \begin{array}{ll}
    x_i + x_j \quad &\mbox{for $x_j \ge x_i$},\\
    x_i + \left( x_j + N \right) \quad &\mbox{for $x_j < x_i$},
   \end{array}\right.\\
  && q = k_1 + k_2, \quad 2k = k_2 - k_1,
\eea
where $k_1$ and $k_2$ denote the individual momenta of the two solitons.
Then the amplitudes $f_{ij}$ can be expressed as follows:
\be
  f_{ij} = \sqrt{\frac{2}{N}} \sum_q {\rm e}^{{\rm i}q \bar{R}} F_q (\bar{r}), \quad F_q (\bar{r}) = \sqrt{\frac{2}{N}} \sum_k {\rm e}^{{\rm i}k \bar{r}} f_q (k),\label{fij}
\ee
where each summation is over $0 \le q, k < \pi$.
This is because both $\bar{R}$ and $\bar{r}$ go by two steps on the lattice spacing.~\cite{Rr}
Substituting eq.\ (\ref{fij}) into eq.\ (\ref{equationfij}), we obtain the equation for $F_q (\bar{r})$:
\bea
  && \left[ E - E_0 - E_{{\rm add}(\bar{r}+1)} \right] F_q (\bar{r}) - \epsilon J(1) \cos q \nonumber \\
  && \times \left[ \left(1 - \delta_{\bar{r},0} \right) F_q (\bar{r}-2) + \left(1 - \delta_{\bar{r},N-2} \right) F_q (\bar{r}+2) \right] = 0.\nonumber \\ \label{Fqr}
\eea
\par

Since we consider the anisotropy near the Ising limit (i.e., $\epsilon \ll 1$), the excitation energy $\omega$ of the Hamiltonian eq.\ (\ref{HamiltonianISD}) is almost $E - E_0$ where $E_0$ is given by eq.\ (\ref{ISIgsenergy}).
Furthermore, we multiply eq.\ (\ref{Fqr}) by $\exp (-{\rm i}k \bar{r})$ and sum on all $\bar{r}$, to obtain the equation for $f_q (k)$:
\be
  \left[ \omega - \tilde{\omega}_q (k ; N) \right] f_q (k) = \frac{2}{N} \sum_{k'(\ne k)} U(k,k') f_q (k'),\label{fk}
\ee
where
\bea
  && \tilde{\omega}_q (k ; N) = E_{\rm c} + 2 \epsilon J(1) \left( 1 - \frac{2}{N} \right) \cos q \cos (2k), \quad \label{IStsc} \\
  && E_{\rm c} = \frac{2}{N} \sum_{\bar{r}} E_{{\rm add}(\bar{r}+1)},\\
  && U(k,k') = \sum_{\bar{r}} {\rm e}^{-{\rm i}(k-k') \bar{r}} E_{{\rm add}(\bar{r}+1)} \nonumber \\
  && \hspace{2cm} - \epsilon J(1) \cos q \left( {\rm e}^{-2{\rm i}k'} + {\rm e}^{2{\rm i}k} \right).
\eea
\par

The scattering states (i.e., two-soliton continuum) are obtained by ignoring the right-hand side of eq.\ (\ref{fk}).~\cite{SS}
The continuum is given by
\be
  \tilde{\omega}_q (k ; \infty) = J \left[ 0.693147 + 2 \epsilon \cos q \cos (2k) \right].\label{continuum}
\ee
Here we have used
\bea
  && J(1) \to J,\\
  && E_{\rm c} \to 0.693147 J,
\eea
for $N \to \infty$.
(The value of $E_{\rm c}$ is the extrapolated one from the numerical calculation for the finite size.)
\par

Let us see the validity of the two-soliton approximation for $N=24$ and $\epsilon=0.1$ (i.e., $\Delta=10$).
In Fig.\ 15(a) we show the result obtained by solving the eigenvalue equation (\ref{fk}) numerically.
This result is compared with the one of the pole positions via the recursion method in Fig.\ 15(b).
This figure shows the poles with main intensity from two-soliton contribution, and exhibits together the ones shifted from $[\pi,2\pi]$ to $[0,\pi]$.
The numbers of the energy levels are the same between the two figures at each momentum $q$, except at $q=5\pi/12, \pi/2$ and $7\pi/12$.
In the two-soliton approximation, all of the degree of freedom appear within the region $0 \le q < \pi$, i.e., half the Brillouin zone.
We judge that this approximation is rather good for $\epsilon=0.1$.
\par

Figure 15(a) suggests the existence of the isolated mode above the continuum spectrum.
In order to make this clear, we show the solutions of eq.\ (\ref{fk}) for $N=100$ and $\epsilon=0.1$ in Fig.\ 16.
First, we verify the good fit by the expression eq.\ (\ref{continuum}) for the continuum spectrum.
Next, one mode separates obviously from the continuum over the whole range of the total momentum $q$.
In addition, another mode is seen for $0.45 \lsim q/\pi \lsim 0.55$.
These isolated modes are interpreted as the sets of the anti-bound states due to the repulsive solitons.
The upper of the two modes corresponds to the isolated mode for the IS $XXZ$ model with $\Delta=10$ obtained by the recursion method.
\par

Here we compare our result with the one for the NN $XXZ$ model by using the same approximation.~\cite{IS}
In this case, the two-soliton continuum is given by
\be
  \omega_q (k,\infty) = J \left[ 1 + 2 \epsilon \cos q \cos (2k) \right],\label{NNtsc}
\ee
with $0 \le q, k < \pi$ in our definition.
No anti-bound state appears when the corresponding eigenvalue equation is solved numerically.
The reason is that the solitons are identified with free particles except for the hard-core repulsion, as mentioned in $\S$\ref{Qualitative}.
\par

The repulsion between solitons turns out to arise from the interaction $J_{ij}$ with $j \ge i+2$ in the Ising term of eq.\ (\ref{HamiltonianISD}).
The necessary interaction for the repulsion between nearest-neighbor solitons is the one which reaches at least the next-nearest-neighbor spin.~\cite{nnncomment}
Independent of the range of the interaction, the terms with $\epsilon$ in the Hamiltonian play two roles: one is to transfer the solitons (i.e., kinetic energy), and the other is to change the number of solitons.
However, the latter role is neglected within the two-soliton approximation.
The kinetic energy causes the dispersion in the excitation spectrum.
\par


\section{Conclusion} \label{Conclusion}

We have investigated the ground-state properties, the energy gap, the DSFs, and the SSFs of the $XXZ$ chain with Ising anisotropy for the two types of the exchange interaction. 
The main results of this study can be summarized as follows:

(1) For the inverse-square type, the overlap between the exact wave function and the Jastrow-type one tends to zero in the thermodynamic limit over the whole range of the Ising anisotropy, except the isotropic case and the Ising limit. 

(2) In each type, the dependence of the energy gap on the anisotropy parameter has the essentially singular form at the critical point.
For the inverse-square type, the opening of the gap is slower than that for the nearest-neighbor type.

(3) With increasing Ising anisotropy, the peak position of $S^{xx}(q,\omega)$ for the nearest-neighbor type approaches the center of the continuum; on the other hand, the peak position for the inverse-square type moves to the upper edge of the continuum, and separates from the continuum for the anisotropy larger than the threshold value.
\par

On the basis of the result (3), we have proposed the picture of the repulsive solitons for the inverse-square $XXZ$ model when the Ising anisotropy is large.
The anti-bound mode in the $(q,\omega)$-plane of $S^{xx}(q,\omega)$ is due to the repulsive solitons.
In a quasi-1D Ising-like antiferromagnet where the exchange reaches a long range, the repulsion will lead to a neutron scattering cross section which is strongly enhanced at the upper boundary or above the continuum.
In this connection, we mention that there are both two-magnon continuum and bound mode for the alternating spin chain.~\cite{TNBGRS}
The existence of the bound mode contrasts with the anti-bound mode in our system.
\par

\section*{Acknowledgments}

We are grateful to H. Yokoyama for instruction of numerical techniques and fruitful discussions, and to B. Sutherland and B. S. Shastry for useful comments.
We acknowledge the use of the diagonalization package, TITPACK Ver.\ 2 by H. Nishimori.
The numerical calculations were performed partly at the Supercomputer Center of the Institute for Solid State Physics, University of Tokyo.
\par

\appendix
\section{Dynamical Structure Factors of the Inverse-Square Ising Model}

The Hamiltonians ${\cal H}_{\rm IS}$ and ${\cal H}_{\rm NN}$ are given in the Ising limit as follows:
\bea
  &&{\cal H}_{\rm IS}/\Delta \stackrel{\Delta \to \infty}{\longrightarrow} {\cal H}_{\rm ISI} = J \sum_{i<j}^N \frac{1}{D(x_i - x_j)^2} S_i^z S_j^z,\label{ISI} \\
  &&{\cal H}_{\rm NN}/\Delta \stackrel{\Delta \to \infty}{\longrightarrow} {\cal H}_{\rm NNI} = J \sum_{i=1}^N S_i^z S_{i+1}^z.\label{NNI}
\eea
In this Appendix, we calculate the DSFs for the IS Ising model defined by eq.\ (\ref{ISI}).
\par

It is not trivial that the ground state for the IS Ising model is the N\'eel state.
To this end, we quote Hubbard's work.~\cite{Hubbard}
He demonstrated that for the 1D model only with the interaction of electrons (negligent of the hopping term), the electrons of the ground state have a periodic arrangement which is regarded as the generalized Wigner lattice under certain conditions.
By adapting his conclusion to the spin system, the following is found:
the exact ground state of an Ising model is the N\'eel state, provided that the interactions $J(r)$ satisfy the two conditions:
\bea
  &&J(r) \to 0 \quad {\rm as} \; r \to \infty,\\
  &&J(r+1) + J(r-1) \ge 2 J(r) \quad {\rm for} \; {\rm all} \; r>1,
\eea
where $r$ denotes the distance between two sites.
The IS Ising model fulfills this conditions for $N \to \infty$.
\par

Next, we calculate the eigenenergies of the N\'eel state and states with two domain walls.
We begin with the N\'eel state.
Let us write the two degenerate N\'eel states as $|{\rm Neel}_1 \rangle \equiv |\uparrow \downarrow \uparrow \downarrow \cdots \uparrow \downarrow \rangle$ and $|{\rm Neel}_2 \rangle \equiv |\downarrow \uparrow \downarrow \uparrow \cdots \downarrow \uparrow \rangle$.
Then
\bea
  {\cal H}_{\rm ISI} |{\rm Neel}_1 \rangle &=& \frac{1}{2} J \left( \frac{\pi}{N} \right)^2 \sum_{i \ne j}^N \frac{1}{\sin^2 [\pi (x_i - x_j)/N]} \nonumber \\
                                       & & \times \frac{1}{4} (-1)^{x_i - x_j} |{\rm Neel}_1 \rangle.
\eea
Using the following relation:
\be
  \sum_{r=1}^{N-1} \frac{(-1)^r}{\sin^2 (\pi r/N)} = - \frac{1}{6} ( N^2 + 2 ),
\ee
the eigenenergy $E_0$ (i.e., the ground-state energy) is given by
\be
  E_0 = - \frac{\pi^2}{48} J N \left( 1 + \frac{2}{N^2} \right).
\ee
The eigenenergy of $|{\rm Neel}_2 \rangle$ is also equal to $E_0$.
Furthermore, we define the state with two domain walls as
\be
  |{\rm DW}(m) \rangle = |\underbrace{\uparrow \downarrow \cdots \uparrow \downarrow \uparrow}_{m} \underbrace{\uparrow \downarrow \cdots \uparrow \downarrow \uparrow}_{N-m} \rangle,
\ee
where $m$ is the distance between two domain walls and should be odd in the subspace $S_{\rm tot}^z = 1$ (or $-1$).
Then the eigenenergy $E_{{\rm DW}(m)}$ of the state $|{\rm DW}(m) \rangle$ is obtained as follows:
\be
  E_{{\rm DW}(m)} = E_0 - \frac{1}{2} J \left( \frac{\pi}{N} \right)^2  \sum_{i=1}^m \sum_{j=m+1}^N \frac{(-1)^{x_i - x_j}}{\sin^2 [\pi (x_i - x_j)/N]}.\label{EDWm}
\ee
Particularly for $m=1$, i.e., the nearest-neighbor domain walls, the double summation of eq.\ (\ref{EDWm}) can be performed to give $E_{{\rm DW}(1)} = E_0 + E_{\rm GDW}$ with
\be
  E_{\rm GDW} = \frac{\pi^2}{12} J \left( 1 + \frac{2}{N^2} \right).
\ee
Also by calculating eq.\ (\ref{EDWm}) numerically, it is found that $E_{{\rm DW}(m)}$ is a monotonically decreasing function of $m$ in the range $1 \le m \le N/2$.
\par

Now that we have found the necessary eigenenergies, it is easy to calculate the DSFs for the IS Ising model. 
If $|{\rm GS}_1 \rangle = (|{\rm Neel}_1 \rangle - |{\rm Neel}_2 \rangle )/\sqrt{2}$ and $|{\rm GS}_2 \rangle = (|{\rm Neel}_1 \rangle + |{\rm Neel}_2 \rangle )/\sqrt{2}$, the longitudinal component $S^{zz}(q,\omega)$ is derived as
\bea
  S^{zz}(q,\omega) &=& \frac{1}{2} \sum_{d=1}^2 \sum_n \left| \langle n | S_q^z | {\rm GS}_d \rangle \right|^2 \delta \left( \omega - ( E_n - E_0 ) \right) \nonumber \\
                   &=& \frac{N}{4} \delta_{q, \pm \pi} \delta(\omega).
\eea
This is the same as the result for the NN Ising model. 
On the other hand, the transverse component $S^{xx}(q,\omega)$ is calculated as follows: 
\bea
  S^{xx}(q,\omega) &=& \frac{1}{2} S^{+-}(q,\omega) \nonumber \\
                   &=& \frac{1}{4} \sum_{d=1}^2 \sum_n \left| \langle n | S_q^- | {\rm GS}_d \rangle \right|^2 \delta \left( \omega - ( E_n - E_0 ) \right) \nonumber \\
                   &=& \frac{1}{4} \sum_{d=1}^2 M_d \delta \left( \omega - ( E_n - E_0 ) \right),\label{SxxqomegaIsing}
\eea
where
\be
  M_d = \frac{1}{N} \sum_n \left| \langle n | \sum_{\ell=1}^N {\rm e}^{- {\rm i} q \ell} S_\ell^- |{\rm Neel}_d \rangle \right|^2.
\ee
The states $|n \rangle$ with nonzero matrix element are limited to those with only one flipped spin from the N\'eel state, i.e., $|{\rm DW}(1) \rangle$ and several of its translationally invariant states; the total number of these states is $N/2$, so that $M_d = 1/2$ ($d=1, 2$).
In addition, all these states have the eigenenergy $E_{{\rm DW}(1)} = E_0 + E_{\rm GDW}$.
As a result, the following equation is obtained:
\be
  S^{xx}(q,\omega) = \frac{1}{4} \delta (\omega - E_{\rm GDW}).
\ee
Incidentally, $S^{xx}(q,\omega) = (1/4) \cdot \delta( \omega - J )$ for the NN Ising model given by eq.\ (\ref{NNI}).
\par

\bigskip



\begin{figure}
\caption{
Ground-state energy per site as a function of $\Delta$.
The thin solid line is the exact result given by eq.\ (\ref{gseanisotropic}) for the NN $XXZ$ model.
The open circles show the data extrapolated from finite size data ($N=6$-$24$) for the IS $XXZ$ model with the second-order polynomial fit.
The value for $\Delta=1$ corresponds to the exact solution given by eq.\ (\ref{HSexactsolution}) of the HS model.
The thick line is guide to the eye.
}
\label{fig:1}
\end{figure}

\begin{figure}
\caption{
The overlap vs the Ising anisotropy $\Delta$ for the IS $XXZ$ model.
$N$ and $M$ denote the numbers of sites and down spins, respectively.
}
\label{fig:2}
\end{figure}

\begin{figure}
\caption{
Size dependence of the overlap for various values of $\Delta$.
}
\label{fig:3}
\end{figure}

\begin{figure}
\caption{
Energy gap as a function of $\Delta$.
The thin solid line is the exact result given by eq.\ (\ref{NNgap}) for the NN $XXZ$ model.
The broken line indicates its asymptote given by eq.\ (\ref{asymptote}) for large $\Delta$.
The open triangles (circles) show the data of $E_{\rm G}^0$ ($E_{\rm G}^\pi$) extrapolated from finite size data ($N=6$-$24$) for the IS $XXZ$ model.
The thick lines are guides to the eye.
}
\label{fig:4}
\end{figure}

\begin{figure}
\caption{
Size dependence of the energy gaps (a) $E_{\rm G}^0$ and (b) $E_{\rm G}^\pi$ for the IS $XXZ$ model with various values of $\Delta$.
The data with each $\Delta$ are fitted by the second-order polynomial.
}
\label{fig:5}
\end{figure}

\begin{figure}
\caption{
$S^{zz}(q,\omega)$ [$=S^{xx}(q,\omega)$] for the IS exchange model with $\Delta=1$ and $N=24$.
The intensity of each pole is proportional to the {\it area} of the circle.
}
\label{fig:6}
\end{figure}

\begin{figure}
\caption{
The $\omega$-dependence of $S^{xx}(\pi,\omega)$ for the IS exchange model with $N=16$-$24$ and $\Delta=1, 2$ and $5$.
The circles, triangles, squares, inverted triangles and diamonds are the data for $N=16, 18, 20, 22$ and $24$, respectively.
These data are obtained by processing of the poles with main intensity (all poles only for $\Delta=1$) via the LSD method.
The inset shows the $\omega$-dependence of $NM^{xx}(\pi,\omega)$ for $\Delta=2$.
}
\label{fig:7}
\end{figure}

\begin{figure}
\caption{
$S^{xx}(q,\omega)$ of the IS exchange model with $N=24$ and (a) $\Delta=2$, (b) $\Delta=5$ and (c) $\Delta=10$.
The intensity of each pole is proportional to the {\it area} of the circle.
}
\label{fig:8}
\end{figure}

\begin{figure}
\caption{
The $\omega$-dependence of $S^{xx}(\pi,\omega)$ for the NN exchange model with $N=16$-$24$ and $\Delta=1, 2$ and $5$.
The symbols are the same as in Fig.\ 7.
These data are obtained by processing of the poles with main intensity via the LSD method.
The inset shows the $\omega$-dependence of $NM^{xx}(\pi,\omega)$ for $\Delta=2$.
}
\label{fig:9}
\end{figure}

\begin{figure}
\caption{
Dynamical quantities for the IS exchange model with $\Delta=10$.
The symbols are the same as in Fig.\ 7.
(a) The $\omega$-dependence of $M^{xx}(\pi,\omega)$ and $NM^{xx}(\pi,\omega)$ (inset) for $N=16$-$24$.
(b) The $\omega$-dependence of $S^{xx}(\pi,\omega)$.
The data are obtained by processing of the poles except highest one via the LSD method.
}
\label{fig:10}
\end{figure}

\begin{figure}
\caption{
The $\omega$-dependence of $M^{xx}(\pi,\omega)$ and $NM^{xx}(\pi,\omega)$ (inset) for the IS exchange model with $N=16$-$24$ and (a) $\Delta=6$ and (b) $\Delta=7$.
The symbols are the same as in Fig.\ 7.
}
\label{fig:11}
\end{figure}

\begin{figure}
\caption{
$S^{zz}(q,\omega)$ of the IS exchange model with $N=24$ and (a) $\Delta=2$ and (b) $\Delta=5$.
The intensity of each pole is proportional to the {\it area} of the circle.
}
\label{fig:12}
\end{figure}

\begin{figure}
\caption{
$S^{xx}(q)$ of the IS exchange model with $N=16$-$24$ and $\Delta=1, 2$ and $5$.
The circles, triangles, squares, inverted triangles and diamonds are the data for $N=16, 18, 20, 22$ and $24$, respectively.
The solid line is the exact result for $\Delta=1$ (i.e., the HS model) given by eq.\ (\ref{sqisotropicl}).
The inset shows the magnification in the region of small $q$.
}
\label{fig:13}
\end{figure}

\begin{figure}
\caption{
$S^{zz}(q)$ of the IS exchange model with $N=16$-$24$ and $\Delta=1, 2$ and 5.
The symbols are the same as in Fig.\ 13.
The solid line is the exact result for $\Delta=1$ (i.e., the HS model) given by eq.\ (\ref{sqisotropicl}).
The inset shows the magnification in the region of small $q$.
}
\label{fig:14}
\end{figure}

\begin{figure}
\caption{
Excitation spectra in the IS exchange model for $N=24$, $\epsilon=0.1$ (i.e., $\Delta=10$) and $S_{\rm tot}^z = 1$.
(a) The closed diamonds show the solutions obtained from the eigenvalue equation (\ref{fk}) using the two-soliton approximation.
The solid lines are the lower and upper edges of the continuum given by eq.\ (\ref{continuum}).
(b) The pole positions by the exact diagonalization and the recursion method [see Fig.\ 8(c)].
The closed diamonds show the raw data in the region $[0,\pi]$, and the open diamonds exhibit the data shifted from $[\pi,2\pi]$ to $[0,\pi]$.
The plus indicates the pole position with very weak intensity ($\sim 10^{-6}$) compared to other ones.
}
\label{fig:15}
\end{figure}

\begin{figure}
\caption{
Excitation spectrum in the IS exchange model for $N=100$, $\epsilon=0.1$ and $S_{\rm tot}^z = 1$.
The closed diamonds show the solutions obtained from the eigenvalue equation (\ref{fk}) using the two-soliton approximation.
The solid lines are the lower and upper edges of the continuum given by eq.\ (\ref{continuum}).
}
\label{fig:16}
\end{figure}


\begin{thebibliography}{99}
\bibitem{Hatsugai} A similar analysis has been performed for the 1D fermionic systems which have the NN hopping and the IS interaction, Y. Hatsugai: 
preprint (cond-mat/9701158).
\bibitem{Rr} Consider the permissible values of $\bar{R}$, which are related to the range of $q$, with $\bar{r}$ fixed. Because both subscripts $i$ and $j$ of $f_{ij}$ permit only even numbers, the values of $\bar{R}$ are limited to even ($2, 4, \cdots, N$) for $\bar{r}/2$ even, and to odd ($1, 3, \cdots, N-1$) for $\bar{r}/2$ odd.
\bibitem{nnncomment} In ref.\ 34, elementary excitations have been studied near the Ising limit in the 1D $XXZ$ model with interactions up to next-nearest neighbors; but repulsive solitons have not been pointed out.
\end{thebibliography}
\end{document}